\documentstyle[11pt,aaspp4,flushrt]{article}

\def\etal{{et al.}\ }
\def\hst{{\it HST\/ }}

\def\hal{H$\alpha$}
\def\hbeta{H$\beta$}
\def\lya{Ly$\alpha$}

\def\nv{\ion{N}{5} $\lambda$1240}
\def\oiv{\ion{O}{4}] $\lambda$1402}
\def\oiv{\ion{O}{4} $\lambda$1402}

\def\civ{\ion{C}{4} $\lambda$1549}
\def\heii{\ion{He}{2} $\lambda$1640}
\def\siiii{\ion{Si}{3}] $\lambda$1882}
\def\ciii{\ion{C}{3}] $\lambda$1909}
\def\nii{\ion{N}{2}] $\lambda$2140}
\def\cii{\ion{C}{2}] $\lambda$2326}

\def\oii{[\ion{O}{2}] $\lambda$2470}
\def\mgii{\ion{Mg}{2} $\lambda$2800}

\def\kms{km s$^{-1}$}

\begin{document}

\title{Ultraviolet Emission from the LINER Nucleus of NGC
6500\footnote{Based on observations made with the NASA/ESA {\it Hubble
Space Telescope}, obtained from the Space Telescope Science Institute,
which is operated by the Association of Universities for Research in
Astronomy, Inc., under NASA contract NAS 5-26555.} }

\author{
Aaron J. Barth\altaffilmark{2}, 
Gail A. Reichert\altaffilmark{3}, 
Luis C. Ho\altaffilmark{4}, 
Joseph C. Shields\altaffilmark{5},
Alexei V. Filippenko\altaffilmark{2}, \nl and
Elizabeth M. Puchnarewicz\altaffilmark{6}
}

\altaffiltext{2}{Department of Astronomy, University of California,
Berkeley, CA 94720-3411.}
\altaffiltext{3}{Code 668, NASA/Goddard Space Flight Center,
Greenbelt, MD 20771.}
\altaffiltext{4}{Harvard-Smithsonian Center for Astrophysics,
Cambridge, MA 02138.}
\altaffiltext{5}{Ohio University, Physics and Astronomy Department,
Athens OH, 45701.}
\altaffiltext{6}{Mullard Space Science Laboratory, University College
London, Holmbury St. Mary, Dorking, Surrey RH5 6NT, UK.}

\begin{abstract}

As a step towards clarifying the ionization mechanism of LINERs, we
have used the {\it Hubble Space Telescope\/} Faint Object Spectrograph
to obtain an ultraviolet spectrum of the nucleus of NGC 6500.  This is
the first time that such a spectrum has been taken of a ``LINER 2''
nucleus-- that is, a LINER lacking broad permitted emission lines like
those in Type 1 Seyfert nuclei.  Compared with more luminous Seyfert
nuclei, the ultraviolet emission-line spectrum of NGC 6500 is
remarkable for its low-excitation character: \ciii, \cii, and \mgii\
are the strongest collisionally-excited lines in the spectrum, and
\civ\ is not detected.  We use the emission-line fluxes to test the
hypothesis, advanced by Dopita \& Sutherland (1995), that the
narrow-line regions of LINERs are primarily excited by fast (150--500
\kms) shocks, rather than photoionized by a central continuum source.
The fast shock models are a poor match to the observed spectrum, as
they predict much stronger emission than is observed for
high-excitation lines such as \civ.  Photoionization by an obscured
nonstellar continuum source, or possibly ionization by slower
($\sim100$ \kms) shocks, are more likely explanations for the
emission-line ratios.  The origin of the ultraviolet continuum is
unclear; its overall spectral shape is reasonably well matched by
spectral evolution models for single-burst populations with ages in
the range 70-100 Myr, but no definite stellar absorption features due
to young stars are present.  Alternately, the ultraviolet continuum
(or some fraction of it) may be scattered radiation from a hidden
active nucleus.  We tentatively detect a broad \ion{He}{2}
$\lambda4686$ emission feature, which may be due to Wolf-Rayet stars,
although the ultraviolet spectrum does not show any clear Wolf-Rayet
signatures.  We also find that NGC 6500 is weakly detected in an
archival ROSAT HRI image with a luminosity of $\sim5 \times 10^{40}$
erg s$^{-1}$ in the 0.1--2.4 keV band, within the normal range for
LINERs.

\end{abstract}

\section{Introduction}

Low-ionization nuclear emission-line regions, or LINERs, occur in
one-third of all bright, nearby galaxies and in nearly 50\% of
early-type (E--Sab) galaxies (\cite{h80}; \cite{hfs5}), yet in most
cases the physical mechanism responsible for the ionization is not
well understood\footnote{As defined by Heckman (1980), a LINER has
[\ion{O}{2}] $\lambda3727$/[\ion{O}{3}] $\lambda5007$ $\geq1$ and
[\ion{O}{1}] $\lambda6300$/[\ion{O}{3}] $\lambda5007$ $\geq\onethird$
.}.  It is likely that LINERs comprise a heterogeneous class in which
stellar and nonstellar photoionization, as well as shocks, contribute
in varying degrees to the energetics of individual objects.  The
presence of broad \hal\ emission in approximately 20\% of LINERs
(\cite{hfs4}) is a clear indication of a link with more luminous
active galactic nuclei (AGNs), and one can define ``LINER 1'' and
``LINER 2'' subclasses, based on the presence or absence of broad
\hal\ emission, in a direct analogy to the Seyfert galaxy population.
The status of the more numerous LINER 2s is less clear, however.  Some
fraction of these objects may be obscured LINER 1 nuclei, or
low-luminosity AGNs which simply lack a luminous broad-line region
(BLR).  Nevertheless, shock excitation and photoionization by young,
massive stars are viable alternatives for the ionization mechanism of
the narrow-line gas in these objects (see Filippenko 1996 for a
review).  If LINER 2 nuclei are indeed a subclass of the AGN family,
then they are the most common form of AGN to occur in nearby galaxies.
It is therefore of great importance to clarify their nature if we are
to develop a complete understanding of the AGN phenomenon at low
luminosities and of the interstellar medium in the central regions of
early-type galaxies.

Recently, Dopita \& Sutherland (1995, 1996; hereafter DS) have
computed new grids of fast (150--500 \kms) shock models, and have
shown that the emission-line spectra of fast shocks can match the
range of optical line ratios observed in Seyfert nuclei and LINERs.
Based on this result, they propose that the narrow-line regions (NLRs)
of {\it all\/} AGNs are primarily shock-excited, rather than
photoionized by the central continuum source.  For a critical
discussion of the DS fast shock models, see Morse \etal (1996).
Optical emission-line diagnostics for Seyferts and LINERs do not yield
a definitive test of the shock hypothesis, but one area in which the
shock models disagree most strongly with standard photoionization
models is in their predictions for the UV spectra of LINERs.  Because
of the high temperatures generated by the passage of a shock front,
the fast shock models predict far stronger emission in high-excitation
UV lines than do traditional photoionization models for LINERs.  Thus
far, however, only three \hst UV spectra of LINERs have been
published: the LINER 1 nuclei of M81 (\cite{hfsm81}) and NGC 4579
(\cite{bar96b}) have UV spectra which are more consistent with the
predictions of photoionization models than with the fast shock models,
while an {\it off-nuclear} spectrum of M87 (\cite{dop96}) reveals
high-excitation emission which may be more consistent with the shock
hypothesis.

NGC 6500 presents a particularly interesting target for such
observations.  This galaxy (type Sab; $cz$=3003 \kms) is a well-known
LINER 2 with strong emission lines (e.g., \cite{fs85}).  Radio
observations reveal the presence of a compact flat-spectrum nuclear
source (\cite{cra79}; \cite{jst81}) as well as extended lobes of
emission emerging perpendicular to the galactic major axis
(\cite{uph89}).  The lobes appear to be the signature of a nuclear
outflow or wind, and the ionized gas shows kinematic signatures of an
outflow as well (\cite{gdp96}).  A diffuse $\sim100$ pc region of UV
emission at the nucleus (\cite{bar96a}) suggests that a nuclear
starburst could be responsible for the outflow, and the possible
presence of Wolf-Rayet (WR) spectral features in the nuclear spectrum,
as we discuss in $\S3.1$ below, reveals that star formation may have
occurred within the last few Myr as well.  In the circumnuclear
region, low-ionization emission is seen out to radii of $\sim1$ kpc
(see the off-nuclear spectra presented by \cite{fil84} and
\cite{gdp96}), unlike the case in the majority of LINERs where the
emission is largely confined to the inner few hundred pc (\cite{p89}).
This evidence for a nuclear outflow makes NGC 6500 an ideal object
with which to test the predictions of shock models for LINERs.  In
this paper, we present \hst UV spectra of the nucleus of NGC 6500, and
apply our results to a test of the fast shock hypothesis.  This is the
first LINER 2 nucleus for which such observations have been obtained.
It has been argued (\cite{dop96}) that the more Seyfert-like LINER 1
nucleus of M81 is not a valid testing ground for the shock models
because of the somewhat higher excitation level of its optical
emission lines, but NGC 6500 is classified as a {\it bona fide} LINER
(\cite{gdp96}; \cite{hfs3}) and such criticisms should not apply to
it.  Throughout this paper we assume a Hubble constant of $H_0=65$
\kms\ Mpc$^{-1}$, corresponding to a distance of 46 Mpc to NGC 6500.

\section{Observations and Measurements}

NGC 6500 was observed by the post-refurbishment \hst Faint Object
Spectrograph (FOS) on 1994 August 13 UT through the 0\farcs86-diameter
circular aperture.  A three-stage peakup with an expected accuracy of
0\farcs12 was used to center the aperture on the optical nucleus of
the galaxy.  Three grating/detector combinations were used, covering
the range 1160--3270 \AA: the G130H grating with the FOS/BLUE detector
(1.0 \AA\ diode$^{-1}$), and the G190H (1.5 \AA\ diode$^{-1}$) and
G270H (2.1 \AA\ diode$^{-1}$) gratings with the FOS/RED detector.
Total exposure times were 10260, 3110, and 1020 s, respectively.  The
point source spectral resolution in each setting is 0.96 diode,
corresponding to 220 \kms, while a source uniformly filling the
aperture would have a spectral resolution of 520 \kms.

The data were processed by the routine \hst calibration pipeline,
which includes flat-fielding, subtraction of the particle-induced
background, and flux and wavelength calibration.  For each grating
setting, we co-added and rebinned the spectra to a wavelength bin size
corresponding to the width of one diode.  The calibrated spectra are
shown in Figure 1.  From the emission-line peaks, we find that small
offsets occur between the wavelength scales of the three gratings (see
\cite{km95} for a discussion of FOS wavelength inaccuracies), and the
spectra in Figure 1 are displayed after shifting to a common
wavelength scale and removing the galaxy's recession velocity.

The FOS scattered light problem (e.g., \cite{ros94}), which primarily
affects the G130H setting, is particularly severe for these data
because of the extremely faint UV continuum of NGC 6500.  By
examination of the uncalibrated G130H data frames in the unilluminated
regions shortward of 1150 \AA, we estimate that $\sim25\%$ of the
total counts at 1300 \AA\ are due to scattered light, while another
$\sim50\%$ are due to particle-induced background light.  The standard
calibration pipeline corrects for these two sources of contamination,
but the scattered light correction is only approximate, as the
distribution of scattered light over the array is not well known.  The
spectral shape of the continuum is drastically altered by the
correction, and we cannot reliably recover the spectral shape below
1600 \AA.  The scattered light correction does not affect the
strengths of emission features, however.

The G190H and G270H spectra have much higher count rates and should
not be strongly affected by scattered light, but in these settings
there are no unilluminated diodes that can be used to estimate the
scattered light contribution.  In the overlap region between the G190H
and G270H settings (observed wavelength 2224--2310 \AA), there is an
upturn in the continuum in the G190H spectrum which does not appear in
the G270H spectrum, and the flux level in G190H is 20\% greater than
in G270H over this wavelength range.  To construct the complete
spectrum, we took the mean of the two settings in this region.  As a
consistency check on the flux scale, we compared our data with the
count rate measured in a WFPC2 F218W image of NGC 6500
(\cite{bar96a}), using the SYNPHOT package within IRAF\footnote{IRAF
is distributed by the National Optical Astronomy Observatories, which
are operated by the Association of Universities for Research in
Astronomy, Inc., under cooperative agreement with the National Science
Foundation.} to measure the F218W magnitude of the FOS spectrum.  We
find that the FOS-derived flux is 33\% greater than the flux level of
the WFPC2 image; the cause of this discrepancy is unknown.

To measure the UV emission-line strengths, we first removed the
continuum around each line by fitting a low-order spline to the local
continuum and subtracting the spline from the data.  Line intensities
were then measured by direct integration of the flux, and, when
possible, by Gaussian fits.  Table 1 lists the measured emission-line
fluxes.  The uncertainty in the placement of the continuum level is a
major source of error for \heii, which lies at the very noisy blue end
of the G190H spectrum.  Neglecting calibration errors, the
uncertainties in the emission-line fluxes are roughly $\sim15\%$ for
the stronger lines but are probably $\sim30-50\%$ for \heii, \siiii,
and \oii.  The line intensities are not corrected for the effects of
interstellar absorption, which is clearly significant for \mgii\ and
possibly also for \lya.  

The full-width at half-maximum (FWHM) was measured by Gaussian
fitting, but meaningful results were obtained only for \lya, \ciii,
and \cii.  Instrumental broadening was subtracted in quadrature from
the line widths using the extended-source resolution of 520 \kms, as
the nuclear emission-line region of NGC 6500 is probably better
approximated by a uniform source than a pointlike source on 1\arcsec\
scales.

Because of the limited signal-to-noise ratio (S/N) of the spectrum, a
few lines are contaminated by noisy diodes or have ambiguous
identifications.  A noisy diode may contribute to the apparent flux of
the \oii\ line; we have attempted to measure the true line flux by
manually truncating the blue half of the line profile to match the red
half, but the resulting flux is probably uncertain by $\sim50\%$.  A
noisy diode also appears at the expected position of \nii\ and renders
impossible the detection of that line.  A weak emission feature of
formal significance $2\sigma$ appears near the expected location of
\nv, but it may simply be a peak in the noise, particularly because
other high-excitation lines such as \civ\ and \oiv\ are not detected
in the G130H spectrum.  We therefore interpret the possible \ion{N}{5}
measurement as an upper limit.  Two emission features are present at
rest wavelengths 2611 and 2626 \AA.  These wavelengths are coincident
with those of two transitions in the UV1 multiplet of \ion{Fe}{2}, but
it is also possible that the features are due to noisy diodes.

The main result of this paper is the non-detection of \civ\ emission.
We determined a $3\sigma$ upper limit to the flux of \ion{C}{4} of
$1.0 \times 10^{-15}$ erg s$^{-1}$ cm$^{-2}$ by examining the
continuum noise in the region 1500--1600 \AA.  The FWHM of \ion{C}{4}
was assumed to be 1200 \kms, since \ion{C}{4} is likely to be at least
as broad as \ciii.

Optical emission-line fluxes measured from starlight-subtracted
spectra have been given by Ho \etal (1997c).  Since the ground-based
spectra were obtained through a 2\arcsec $\times$ 4\arcsec\ effective
aperture, we cannot directly compare the UV and optical emission-line
strengths.

NGC 6500 lies at a Galactic latitude of 20\arcdeg, and its spectrum is
affected by a substantial amount of extinction.  Ho \etal (1997c)
measure a Balmer decrement of \hal/\hbeta = 3.64 for NGC
6500. Assuming an intrinsic Balmer decrement of 3.1, which is a
typical value for the shock and photoionization models discussed
below, we estimate a total reddening of $E(B-V)$ = 0.16 mag for the
nucleus of NGC 6500.  The Galactic \ion{H}{1} column toward NGC 6500
is $7.0\times10^{20}$ cm$^{-2}$ (\cite{mur96}).  Using the conversion
factor given by Mathis (1990), this \ion{H}{1} column implies that the
Galactic contribution to $E(B-V)$ is 0.12 mag, somewhat higher than
the estimate of $E(B-V) = 0.09$ mag given by Burstein \& Heiles
(1984).  Extinction-corrected line intensities, computed using the
extinction curve of Cardelli \etal (1989), are listed in Table 1.  It
is important to bear in mind that the reddening curve within the NGC
6500 nucleus may be very different from the Galactic law, but Galactic
dust is the dominant contribution to the reddening of this object, and
in any case the reddening correction has less than a 10\% effect on
our primary diagnostic of the NLR ionization level, the
\ion{C}{3}]/\ion{C}{4} flux ratio.

An archival ROSAT HRI (0.1--2.4 keV) image taken on date 1993 March 17
shows that NGC 6500 is detected at the $3\sigma$ level in a 5954 s
exposure, with a total of $15\pm5$ counts above background within a
square aperture of size $24\arcsec\times24\arcsec$.  Using the
energy-to-count rate conversion data in the ROSAT HRI Calibration
Report (\cite{dav96}), and assuming an $f_{\nu} \propto \nu^{-1}$
continuum and the Galactic \ion{H}{1} column given by Murphy \etal
(1996), we find an unabsorbed source flux of $2 \times 10^{-13}$ erg
cm$^{-2}$ s$^{-1}$.  This flux corresponds to an X-ray luminosity of
$5 \times 10^{40}$ erg s$^{-1}$ in the 0.1--2.4 keV band, which is
within the normal range for LINERs (see \cite{fab96} and \cite{spy96})
and superwind galaxies (\cite{ham90}).

\section{Discussion}

\subsection{The UV Continuum}

The region emitting the UV continuum is diffuse, with a diameter of
$\sim100$ pc, and it lacks a central peak at the location of the
optical nucleus (\cite{bar96a}).  By subtracting model point-spread
functions from the WFPC2 image, we estimate that at most $\sim7\%$ of
the 2200 \AA\ flux could be due to a single point source.  Thus, an
unobscured AGN could make only a very small contribution to the UV
continuum.  The two most likely sources of the extended UV emission
are starlight and scattered radiation from a hidden AGN, and we
discuss these possibilities below.

Correction of the continuum shape for the total reddening of
$E(B-V)=0.16$ mag using a Galactic extinction law leads to an
unacceptably large bump at 2200 \AA\ in the corrected spectrum.  Even
correcting for $E(B-V)=0.12$ mag, the extinction inferred from the
Galactic \ion{H}{1} column, leads to a distinct rise in the continuum
at 2200 \AA.  Using the Burstein \& Heiles (1984) value of
$E(B-V)=0.09$ mag does not result in a noticeable 2200 \AA\ bump.  The
dangers of applying a Galactic extinction law to extragalactic
starburst regions have been discussed by Calzetti \etal (1994), who
conclude that the internal extinction law within starbursts is flatter
than the Galactic law and lacks the 2200 \AA\ feature, and also that
the optical continua of starbursts are less reddened than the emission
lines.  Since it is not clear how to best correct the UV continuum of
NGC 6500 for the effects of extinction, particularly for the internal
component, we choose to correct the UV continuum only for
$E(B-V)=0.09$ mag using the Galactic extinction law, and we warn that
we may have not fully removed the effects of extinction from the
continuum.
 
An upturn in the continuum longward of 2800 \AA\ can plausibly be
attributed to the increasing contribution of late-type stars at these
wavelengths.  An absorption feature at the expected location of the
3096 \AA\ blend adds support to this interpretation, as this feature
(a blend of Fe I lines and \ion{Al}{1} $\lambda3093$) appears in
stellar spectra of type F7 and later (\cite{fan90}).  In the remainder
of this section we address the question of the origin of the continuum
flux shortward of 2800 \AA. 

Due to the low S/N in the continuum, there are no definite stellar
absorption features that can be identified with known features in
stellar UV spectra.  A broad depression of equivalent width (EW)
$\sim4$ \AA\ appears in the region 1536--1545 \AA\ (rest wavelength),
and may be blueshifted \ion{C}{4} stellar-wind absorption.  A spectrum
with higher S/N would be necessary to confirm the reality of this
feature, and we interpret the measured EW of 4 \AA\ as an upper limit
rather than a detection.  We place an upper limit of 2 \AA\ on the EW
of any \ion{Si}{4} $\lambda1400$ absorption.  The only definite
absorption features in the spectrum are the Galactic and internal
interstellar absorption features due to the UV2 multiplet of
\ion{Fe}{2} (2586 and 2599 \AA) and the \ion{Mg}{2} doublet, which
appear in the G270H spectrum.

The shape of the UV continuum is roughly similar to those of archival
{\it International Ultraviolet Explorer (IUE)} spectra of late B-type
and early A-type stars, suggesting that the UV continuum may originate
in a young stellar population.  To determine the age of the stellar
population that most closely matches the observed spectrum, we compare
the data with synthetic spectra generated by the isochrone synthesis
code GISSEL96 (\cite{bc93}).  Figure 2 shows the UV continuum shape of
NGC 6500 and the spectra predicted for single-burst models of solar
metallicity and a Salpeter (1955) initial mass function (IMF), for
three different burst ages.  The overall spectral shape is bracketed
by burst ages ranging from 36 Myr to 130 Myr, with the best overall
match to the data for ages of 70--100 Myr.  Because of the large
uncertainty in the continuum shape below 1600 \AA, there is little
information in the spectrum that can distinguish the single-burst
models from models with more complex star-formation histories, and we
restrict our attention to the single-burst models.  The model spectra
with $t\gtrsim130$ Myr are clearly too red to match the data, allowing
us to set an upper age limit of 130 Myr for the stellar population
responsible for the bulk of the UV emission.  In the wavelength range
1300--2300 \AA, the 36 Myr burst has a slope roughly similar to that
of NGC 6500, and because of the uncertainties in the reddening and in
the continuum shape below 1600 \AA, we cannot rule out a substantial
contribution from even younger stars.  No single-burst model can fit
the entire UV spectral region, however, as the upturn longward of 2800
\AA\ indicates the presence of an older population of stars.  An
excess over the model spectra is present at 2300--2500 \AA, partly due
to a few sharp ``bumps'' in the region between \cii\ and \oii, and we
are unable to explain this excess in terms of any normal stellar
population.

To what extent might this young stellar population be responsible for
the NLR excitation?  Shields (1992; see also \cite{ft92}) has shown
that early O-type stars can give rise to a LINER-like optical spectrum
if the surrounding interstellar medium is sufficiently dense.  The
possible presence of WR emission in the optical spectrum (see below)
indicates that stars may have been forming within the last few Myr,
and the possible blueshifted \ion{C}{4} absorption may be a further
indication of stars with $M>30 M_{\sun}$ (\cite{lrh95}).  As a rough
comparison with our upper limits to the strength of \ion{C}{4} and
\ion{Si}{4} absorption in NGC 6500 (4 \AA\ and 2 \AA, respectively),
we note that the brightest starburst knot in NGC 4214, which has an
age of 4--5 Myr, has EW(\ion{C}{4}) = 5.9 \AA\ and EW(\ion{Si}{4}) =
3.9 \AA\ (\cite{lei96}). Thus, if a similarly young population were
present in NGC 6500, its 1500 \AA\ continuum would likely be diluted
by an equal or greater contribution from B and A stars, or by
scattered AGN continuum radiation.

If the NLR luminosity were {\it solely} due to ionization by massive
stars in the nucleus, then the ionizing photon luminosity of the
nuclear starburst region must be at least $2.7\times10^{52}$ s$^{-1}$
to account for the nuclear \hbeta\ flux, assuming Case B
recombination.  To obtain a hard upper limit to the possible ionizing
photon luminosity due to stars in the nucleus, we assume that the 1500
\AA\ continuum flux arises entirely from a young starburst with age
$<5$ Myr, and we apply the full extinction correction of $E(B-V)=0.16$
mag to this flux.  Using the results of Leitherer \& Heckman (1995)
and Leitherer \etal (1995) for a solar-metallicity instantaneous burst
with a power-law IMF of slope --2.35 and upper mass cutoff 100
$M_{\sun}$, we find that the extinction-corrected 1500 \AA\ flux
translates to a population of $\sim10^3$ O-type stars and an ionizing
photon luminosity of $2.0\times10^{52}$ s$^{-1}$, or 74\% of the
amount required to ionize the NLR.  However, the true ionizing photon
output from hot stars is likely to be far less than this figure,
because the apparent weakness of \ion{C}{4} and \ion{Si}{4} absorption
imply that the majority of the 1500 \AA\ flux is not likely to come
from O-type stars, and because the extinction correction of
$E(B-V)=0.16$ mag is likely to be an overcorrection.  We cannot rule
out a sizeable contribution by hot stars to the NLR energetics, but it
seems unlikely that hot stars could be the dominant contribution.

The possible detection of an optical \ion{He}{2} $\lambda4686$ feature
provides supporting evidence for a recent burst of star formation.
Figure 3 shows the starlight-subtracted optical spectrum of NGC 6500
(from Ho \etal 1997c), illustrating the 4600--4700 \AA\ feature, which
we tentatively attribute to WR stars.  Its profile appears similar to
the WR feature found in NGC 3049 (\cite{vc92}): \ion{He}{2}
$\lambda4686$ is clearly present, nebular [\ion{Fe}{3}] $\lambda4658$
emission is superposed on the WR emission complex, and \ion{C}{4}
$\lambda4658$ from WC stars may be present as well.  There may be weak
emission from \ion{N}{3} $\lambda4640$, but the feature at this
wavelength does not stand out above the noise level.  NGC 6500 differs
from most WR galaxies in that its nuclear optical continuum is
dominated by old stars; most known WR galaxies have a relatively
featureless, blue optical continuum due to massive stars.  In NGC
6500, the WR emission blend only becomes visible after subtraction of
the underlying starlight continuum.  The width of the \ion{He}{2}
emission line adds support to the WR interpretation: a single-Gaussian
fit to \ion{He}{2} yields FWHM $\approx1200$ \kms, while the strong
optical emission lines have FWHM = 550--700 \kms\ (\cite{hfs3}).
Taking care not to include the adjacent [\ion{Fe}{3}] $\lambda4658$
emission, we find a flux of $2.5\times10^{-15}$ erg s$^{-1}$ cm$^{-2}$
for the dereddened \ion{He}{2} line.  Compared with the noise level in
the region 3280--4630 \AA, the significance of the \ion{He}{2}
detection is $2.6\sigma$.  The true significance of the detection must
be lower than this value, because of the unknown systematic
uncertainties involved in the starlight subtraction as well as the
possible nebular contribution to the $\lambda4686$ feature.

Using the calibrations given by Vacca \& Conti (1992), the strength of
the \ion{He}{2} $\lambda4686$ WR emission (less the unknown nebular
contribution) implies the presence of $<350$ WN stars.  Depending on
the age and metal abundance of the young stellar population, the WR/O
star ratio can vary from $\lesssim0.1$ to $>1$ (\cite{lh95}), so the
WR census is at least consistent with the upper limit on the O-star
population given above.  Optical spectra from the Space Telescope
Imaging Spectrograph (STIS) could resolve much of the ambiguity in
these estimates, as the narrow STIS slit would exclude most of the
surrounding bulge starlight, enabling a more accurate measurement of
the WR emission blend and a search for other, weaker WR spectral
features.

In addition to the possible young stellar population, there may be an
accretion-powered active nucleus in NGC 6500.  If we allow for the
possibility that the emission lines in NGC 6500 result from
photoionization by an AGN-like continuum source, then we can estimate
the UV flux that would be received from the central source in the
absence of any obscuration.  Assuming a power-law continuum with
$f_\nu \propto \nu^{-\alpha}$, the slope $\alpha$ and the \hbeta\ EW
are related by EW(\hbeta) = ($560/\alpha$)(3/16)$^\alpha$ \AA\ in
the case of complete covering of the source by clouds optically thick
to Lyman continuum photons (\cite{fn83}).  The ionizing continuum need
not be a single power law, particularly if a ``big blue bump'' is
present, but the spectral energy distribution of M81 shows that at
least some LINERs lack this feature (\cite{hfsm81}).  In the more
realistic case of incomplete source covering, this relation provides
an upper limit to the \hbeta\ EW.  Using the \hbeta\ flux from Ho
\etal (1997c) and assuming $\alpha=1$, we find that the predicted
continuum flux from a central AGN, if it were emitted isotropically
and completely unobscured except for Galactic extinction, would be
$\geq6\times10^{-16}$ erg s$^{-1}$ cm$^{-2}$ \AA$^{-1}$ at 2500 \AA.
This limit is a factor of 2.4 greater than the observed 2500 \AA\
continuum flux, indicating that if the LINER emission in NGC 6500 were
solely due to nonstellar photoionization, then the central source must
be heavily obscured.  With the constraint that a point source makes up
at most 7\% of the UV continuum flux, the internal extinction toward
the central continuum source would be at least $E(B-V)=0.58$ mag.
Given this evidence for an ``ionizing photon deficit,'' it is tempting
to speculate that there may be a nonstellar UV continuum which is
emitted anisotropically, or obscured by a dense molecular torus, as is
presumed to occur in many Seyfert 2 galaxies (\cite{wwh88}).

If NGC 6500 does contain an AGN hidden within an obscuring torus, then
scattered AGN radiation could be an alternative source of the
spatially extended UV continuum emission.  WFPC2 UV imaging
polarimetry could reveal whether scattered radiation contributes
substantially to the UV continuum.  One possible argument against a
scattering origin for the bulk of the UV continuum emission comes from
the fact that no broad lines are detected in NGC 6500.  In the LINER 1
nuclei of M81 and NGC 4579, the broad \ion{C}{4} emission has an EW of
140 and 150 \AA, respectively (\cite{hfsm81}; \cite{bar96b}).  If we
suppose that the central engine of NGC 6500 is simply an obscured
version of those in M81 and NGC 4579, and if the scattering geometries
are the same for the continuum source and BLR, then broad \ion{C}{4}
should be seen in the extended emission, with an EW of $\sim150$ \AA\
with respect to the scattered continuum.  A conservative upper limit
to the EW of broad \ion{C}{4} in NGC 6500 is 10 \AA, where we have
assumed FWHM = 4700 \kms, the average of the broad \ion{C}{4} widths
of NGC 4579 and M81.  However, this argument rests on the assumption
that the intrinsic EW of the broad-line emission in NGC 6500 is
similar to those of the other LINER 1 galaxies, which need not be the
case.

To summarize, we find that the UV continuum shape is reasonably well
represented by a starburst of age 70-100 Myr, although older stars
clearly contribute to the continuum longward of 2800 \AA.  A
substantial contribution to the far-UV continuum from even younger
stars cannot be ruled out, particularly given the uncertainty in the
extinction correction, but we do not detect definite UV spectral
features from WR or O stars.  The possible \ion{He}{2} $\lambda4686$
feature is the only evidence for such a young population.  It is
conceivable that a population of WR and early O-type stars could be
partly responsible for the LINER emission, although it is unlikely
that hot stars alone could account for the majority of the NLR
luminosity.  If the NLR is photoionized by a nonstellar continuum,
then the central source must be obscured along our line of sight; a
search for ionization cones with \hst narrow-band images could test
this possibility.  Scattered AGN radiation could contribute to the
spatially extended UV continuum emission, but without polarimetric
observations there is no clear way to estimate what fraction of the UV
continuum might be nonstellar in origin.

\subsection{Emission-Line Diagnostics}

Examination of the UV spectrum shows that the NLR of NGC 6500 is
remarkable for its low-excitation characteristics.  In contrast to
low-luminosity Seyfert nuclei such as NGC 1566 (\cite{kri91}) and NGC
4395 (\cite{fhs93}), the low-ionization lines of \cii\ and \mgii\ are
among the strongest in the UV spectrum of NGC 6500, while \civ\ is
undetected.  The UV emission lines, with FWHM = 730--1050 \kms, are
broader than the optical lines, which have FWHM of 550--700 \kms, but
this discrepancy may be solely due to the difference in aperture sizes
between the UV and optical measurements.  The UV line profiles do not
show broad wings, and we conclude that they arise from the NLR and
that no BLR is detected.

Applying the IRAF task TEMDEN to the emission-line fluxes measured by
Ho \etal (1997c), we have estimated the density and electron
temperature of the NLR with standard optical line-ratio diagnostics.
For the [\ion{S}{2}] lines, we find the ratio $I(6717)/I(6731) =
1.13$, indicating a density of $n_e = 350$ cm$^{-3}$ (for $T_e=10^4$
K) or $n_e = 460$ cm$^{-3}$ (for $T_e=2.5\times10^4$ K) for the
[\ion{S}{2}]-emitting region.  From the [\ion{O}{3}]
$\lambda\lambda4959,5007 / \lambda4363$ ratio of 27.9, we find $T_e =
2.6 \times 10^4, 2.1 \times 10^4$, and $1.1 \times 10^4$ K for
densities of $n_e = 10^2, 10^5$, and $10^6$ cm$^{-3}$, respectively.
Such high inferred temperatures indicate that either high-density
($\gtrsim 10^6$ cm$^{-3}$) regions are present in the NLR, or,
alternatively, that shock heating plays a significant role.

Photoionization models with low ionization parameter make quite
different predictions from the DS fast shock models for the strengths
and ratios of the UV emission lines.  In the shock models, the cooling
postshock region radiates strongly in lines of highly ionized species
because of the high electron temperatures generated by the shock.  The
primary difference between LINER and Seyfert spectra, in the DS model,
is that LINERs lack sufficient gas to form precursors, the preshock
regions photoionized by extreme-UV and X-ray photons generated in the
shock.  Without such precursor regions, the shock models predict a
low-excitation optical spectrum similar to those of LINERs, and a
high-excitation UV spectrum in which lines such as \civ\ and \nv\ are
important coolants.  On the other hand, in standard photoionization
models with $U \approx 10^{-3.5}-10^{-4.0}$, which roughly reproduce
the optical emission-line spectra of LINERs, lines of highly ionized
species such as \ion{C}{4} and \ion{N}{5} are extremely weak or
totally absent because of the low ionization parameter.  Thus, the
strength of the high-excitation UV lines is an important diagnostic
that can help to discriminate between the two classes of models.

To investigate the ionization level of the NLR gas, we compare the
relative strengths of the UV lines of carbon: \cii, \ciii, and \civ.
In Figure 4 we plot the measured ratio (\ion{C}{3}]/\ion{C}{4})
against (\ion{C}{2}]/\ion{C}{3}]) and the values predicted by various
models.  Three sets of model results are included; we describe the
basic model parameters here, and refer the reader to the sources
listed below for full details on the model calculations.  For the DS
fast shock calculations, a preshock density of $n$(H) = 1 cm$^{-3}$ is
assumed, the shock velocity is in the range 150--500 \kms, and model
results are plotted for two values of the ``magnetic parameter''
$B/n^{1/2}$, which controls the compression of the postshock gas.  For
slower shocks, we include the results of models C--G from Raymond
(1979), with velocities of 70--140 \kms\ and preshock density of
$n$(H) = 10 cm$^{-3}$.  The photoionization model results are adapted
from calculations performed by Ho \etal (1993) using the
photoionization code CLOUDY (Ferland 1991).  These calculations were
done for a single slab of gas illuminated by a power-law continuum
with $f_\nu \propto \nu^{-1.5}$, and we display the results for
densities ranging from $n_e=10^{2.5}$ to $10^{6.0}$ cm$^{-3}$ and
ionization parameters ranging from log $U = -2.5$ to $-3.5$.  All of
the models used an undepleted solar abundance set (e.g., \cite{ga89}),
except that Ca and Fe were each depleted by 0.2 dex in the DS
calculations.

The differences between the DS shock model predictions and
photoionization calculations are readily apparent in this diagram, as
the DS models all predict \ion{C}{4} $>$ \ion{C}{3}], while all
photoionization models with log $U \leq -2.5$ have \ion{C}{3}] $>$
\ion{C}{4}.  In the DS models, the \ion{C}{3}]/\ion{C}{4} ratio ranges
from 0.1 to 0.9, but even in the most favorable case of 500 \kms\
shocks with high magnetic parameter, the predicted ratio is a factor
of 4.5 too small to be consistent with the observed limit.  Other line
ratios are in conflict with the data as well: the fast shock models
predict that the \nv\ and \oiv\ lines should be comparable to or
stronger than \ion{C}{3}], and that for low magnetic parameter and low
shock velocities ($\leq200$ \kms), \ion{O}{3}] $\lambda1666$ should be
comparable to or stronger than \ion{C}{3}] and \ion{C}{2}].  The
predictions of the DS fast shock models are clearly at odds with the
low-ionization state of the NGC 6500 NLR.  It is worth noting as well
that the UV emission-line ratios in NGC 6500 are quite unlike those
seen in supernova remnants.  Spectra of shock-excited filaments in the
Cygnus Loop and of Puppis A with shock velocities of $\sim170$ \kms\
show \ion{C}{4} $>$ \ion{He}{2} (\cite{bla91}; \cite{bla95}), while in
NGC 6500 our upper limit on \ion{C}{4} is a factor of 2 weaker than
the strength of \ion{He}{2}.

Slower shocks, on the other hand, might provide a better match to the
observed line ratios.  The Raymond (1979) calculations span a wide
range in the \ion{C}{3}]/\ion{C}{4} ratio, and at the lowest
velocities ($<100$ kms) are consistent with the \ion{C}{3}]/\ion{C}{4}
limit and nearly consistent with the \ion{C}{2}]/\ion{C}{3}] ratio in
NGC 6500.  These models cannot fully describe the NLR conditions
within NGC 6500, though, because they underpredict [\ion{O}{1}]
$\lambda6300$ by an order of magnitude, yielding a spectrum that does
not meet the definition of a LINER.  However, given the tremendous
parameter space available for shock models, it seems likely that some
combination of shock parameters, or some superposition of different
shock models, could adequately reproduce the range of emission-line
strengths observed in NGC 6500.  Certainly, shocks must be occurring
to some extent within the nucleus of NGC 6500, given the kinematic and
morphological evidence for a nuclear outflow.

The lack of \civ\ emission in NGC 6500 is consistent with the
predictions of photoionization calculations as long as log $U \lesssim
-2.5$, and the \ion{C}{2}]/\ion{C}{3}] ratio suggests log $U = -3.0$.
This result is inconsistent with the optical spectrum, however, as the
optical line ratios [\ion{O}{2}]
$\lambda\lambda3726,3729$/[\ion{O}{3}] $\lambda5007$ and [\ion{O}{1}]
$\lambda6300$/[\ion{O}{3}] $\lambda5007$ in NGC 6500 are better fit by
the models with log $U = -3.5$ to $-4.0$.  The mismatch indicates that
a range of values of $U$ must be present if the NLR gas is
photoionized.  The high [\ion{O}{3}] temperature also suggests that
the NLR of NGC 6500 may be stratified in density (\cite{fh84}), and it
is clear that both density and ionization stratification must be taken
into account when constructing realistic photoionization models for
LINERs.  As in the case of NGC 4579 (Barth \etal 1996b), the power-law
photoionization models overpredict the strength of \heii\ (relative to
\ion{C}{3}]) for densities less than $10^6$ cm$^{-3}$, and the
discrepancy is more severe if part or all of the $\lambda1640$ line is
generated in WR atmospheres.  The weak \ion{He}{2} line relative to
photoionization predictions is a likely indication that the ionizing
continuum falls off steeply at energies higher than 54.4 eV (e.g.,
\cite{p84}).  Alternately, the weakness of \ion{He}{2} in LINERs could
result from a composite cloud population consisting of optically thin
and optically thick components, as in the scenario of Binette \etal
(1996).

It would be premature, however, to conclude that the NLR of NGC 6500
is entirely photoionized by an AGN, and it is not clear that this
ambiguity can be resolved with the present data, especially in view of
the overlap in Figure 4 between the photoionization models and the
slow shock models.  If the spatially extended emission arises from gas
illuminated by a central source, then the roughly constant optical
emission-line ratios suggest that the ionization parameter does not
vary strongly with radius, which would imply that $n_e \propto r^{-2}$
(\cite{fil84}).  However, the spatially extended low-ionization
emission may also be suggestive of shock excitation, or possibly of a
spatially distributed source of ionizing photons.  In addition, it is
well known that the extended emission in superwind galaxies can have a
LINER-like optical spectrum (\cite{ham87}).  Given the evidence for a
nuclear outflow in NGC 6500, it seems at least plausible that the
extended NLR emission could be generated by shocks in the expanding
wind, while the nuclear emission might be due to gas photoionized by a
hidden AGN.  Long-slit spectra taken with STIS could be used to search
for spatial gradients in the emission-line ratios that might indicate
a transition from photoionization at the nucleus to shock excitation
in the extended emission.  A comparison of the UV emission-line
spectrum with the predictions of composite shock plus photoionization
models (e.g., \cite{con97}) would be instructive.  As discussed in \S
3.1, hot stars at the nucleus may also contribute a sizeable fraction
of the ionizing photon luminosity. Deep STIS UV spectra of NGC 6500
would also enable a more sensitive search for \ion{C}{4} emission and
a better assessment of the importance of hot stars in this object, as
well as an independent check on the reality of the possible \ion{N}{5}
emission and \ion{C}{4} absorption.

At a more fundamental level, the key question that must be addressed
is whether LINER 2 nuclei are powered by accretion onto a supermassive
black hole, or by massive stars, or by some other process.  The
ionization state of the NLR gas provides clues but does not yield a
compelling answer to this question.  Is there any evidence that {\it
requires} an AGN to be present in the nucleus of NGC 6500?  Broad
emission lines, one of the clearest signs of such activity, are
conspicuously absent from the UV spectrum, and we can only speculate
as to whether they may be hidden behind obscuring material or whether
they are not present at all in NGC 6500.  If a compact UV continuum
source is present and luminous enough to contribute significantly to
the NLR ionization, then it must be highly obscured.  Optical
spectropolarimetry could reveal hidden broad-line emission and
continuum polarization, and a hard X-ray continuum might be detectable
with AXAF observations.  At present, the compact, flat-spectrum radio
source is perhaps the most direct indication of a hidden AGN in this
galaxy.  As optical and UV studies have been inconclusive, future
radio and X-ray observations may provide the best means for
determining the nature of this object.

\section{Summary and Conclusions}

Optical studies of LINERs generally focus on the problem of
determining whether shock heating or photoionization is the dominant
ionization mechanism within the NLR.  In NGC 6500, we have an example
of a LINER in which shocks, photoionization by an AGN, and ionization
by hot stars may all make energetically significant contributions, and
we hope to reach an understanding of the relationships between the
nuclear starburst, the outflow, and the hidden active nucleus that may
be present.  The available data do not suffice to disentangle these
components from one another, but our study of the UV spectrum of NGC
6500 does lead to the following conclusions:

1. If the UV continuum is due to starlight, then stars of age
$\lesssim100$ Myr are present in the nucleus. The \ion{He}{2}
$\lambda4686$ emission line, which we tentatively attribute to WR
stars, indicates that the region may have experienced a burst of star
formation within the last several Myr.  No clear signatures of such
young stars appear in the UV spectrum, but data of higher S/N may
reveal the P Cygni features of \ion{C}{4} and \ion{Si}{4} which are
expected from such a young population.  Rough order-of-magnitude
estimates indicate that this young burst population may contribute
substantially to the ionization of the NLR, but better data are needed
to fully assess this possibility.

2. If the NLR of NGC 6500 is photoionized by an AGN, then the central
continuum source must be highly obscured along our line of sight but
visible to the NLR.  Future polarization measurements can test whether
scattered radiation from a hidden AGN contributes to the spatially
extended UV continuum emission.

3. The UV emission-line ratios are incompatible with the predictions
of the Dopita \& Sutherland (1995, 1996) fast shock models, which
predict strong lines of \ion{C}{4}, \ion{O}{4}, and other highly
ionized species.  Slower ($\lesssim100$ \kms) shocks can match the UV
emission-line ratios more successfully, as can power-law
photoionization models.

Although the status of NGC 6500 remains somewhat ambiguous, the UV
spectra of the LINERs M81, NGC 4579, and NGC 6500 provide ample
evidence to contradict the claim of DS that the narrow-line regions of
all Seyferts and LINERs are primarily excited by fast shocks.  Other
objects, such as the nuclear disk of M87 (\cite{dop96}), may well be
excited by fast shocks, but there are not yet any examples known of
typical LINER {\it nuclei} having high-excitation UV emission-line
spectra.  Future STIS observations should provide far better data with
which to assess the relative contributions of shocks, hot stars, and
nonstellar continuum sources to the ionization of LINERs.

\acknowledgements

We are grateful to Mike Eracleous, Nancy Levenson, Daniel Stern, Jules
Halpern, and John Raymond for helpful advice, and to an anonymous
referee for very constructive suggestions that improved this paper.
We also owe thanks to Gary Ferland, and to Gustavo Bruzual and
St\'ephane Charlot, for making their excellent software (CLOUDY
and GISSEL96, respectively) available to the community.  This research
has made use of the NASA/IPAC Extragalactic Database (NED), which is
operated by the Jet Propulsion Laboratory, California Institute of
Technology, under contract with NASA.  Financial support was provided
by grants GO-5431-93A, AR-5291-93A, AR-5792-94A, and GO-6112-94A from
the Space Telescope Science Institute, which is operated by AURA,
Inc., under contract with NASA.

\clearpage

\begin{deluxetable}{lrrcc}
\tablewidth{5in}
\tablenum{1}
\tablecaption{Ultraviolet Emission Lines in NGC 6500.}
\tablehead{\colhead{Line} & \colhead{$F\times10^{15}$} &
\colhead{$I\times10^{15}$}  &
\colhead{FWHM} & \colhead{Notes} \nl
\colhead{} & \colhead{(erg cm$^{-2}$ s$^{-1}$)} & 
\colhead{(erg cm$^{-2}$ s$^{-1}$)}  & (km s$^{-1}$) & }
\startdata
Ly$\alpha$ $\lambda$1216 & 29.0\phn\phn\phn\phn\phn\phn & 146\phn\phn\phn\phn\phn\phn   & 730          & a? \nl
N V $\lambda$1240?       &$<1.2$\phn\phn\phn\phn\phn\phn&$<5.7$\phn\phn\phn\phn\phn\phn & \nodata      & b,c? \nl
C IV $\lambda$1549       &$<1.0$\phn\phn\phn\phn\phn\phn& $<3.3$\phn\phn\phn\phn\phn\phn& \nodata      &  \nl
He II $\lambda$1640      & 1.9\phn\phn\phn\phn\phn\phn  & 6.1\phn\phn\phn\phn\phn\phn   & \nodata      &  \nl
Si III] $\lambda$1883    & 1.1\phn\phn\phn\phn\phn\phn  & 3.6\phn\phn\phn\phn\phn\phn   & \nodata      & \nl
C III] $\lambda$1909     & 4.0\phn\phn\phn\phn\phn\phn  & 13.4\phn\phn\phn\phn\phn\phn  & 1050         & \nl
C II] $\lambda$2326      & 4.1\phn\phn\phn\phn\phn\phn  & 14.7\phn\phn\phn\phn\phn\phn  & 820          & \nl
[O II] $\lambda$2470     & 0.7\phn\phn\phn\phn\phn\phn  & 2.1\phn\phn\phn\phn\phn\phn   & \nodata      & c \nl
Mg II $\lambda$2800      & 4.5\phn\phn\phn\phn\phn\phn  & 11.0\phn\phn\phn\phn\phn\phn  & \nodata      & a \nl
\enddata
\tablecomments{$F$ = measured flux; $I$ = flux corrected
for total reddening of $E(B-V)=0.16$ mag.  Extinction correction factors
were computed using the extinction curve of Cardelli et
al. (1989). (a) Affected by absorption. (b) Line identification
uncertain.  (c) Line contaminated by noisy diode.}
\end{deluxetable}

\clearpage
\begin{figure}
\plotone{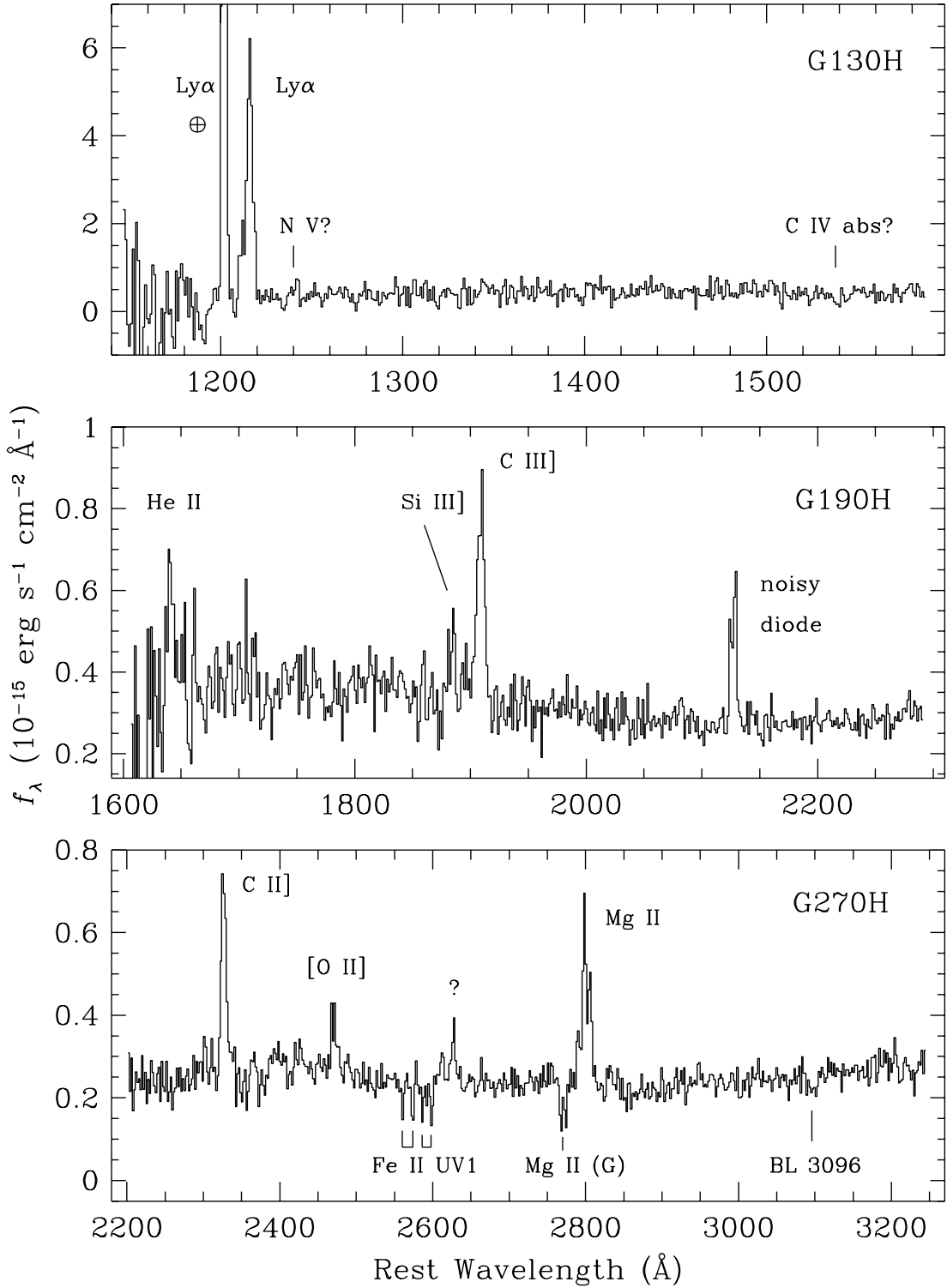}
\caption{Ultraviolet spectrum of NGC 6500.  Emission and
absorption features discussed in the text are labelled.  The
absorption lines labelled as being members of the \ion{Fe}{2} UV1
multiplet include the Galactic and the internal lines at rest
wavelengths of 2586 and 2599 \AA.  Internal absorption due to
\ion{Mg}{2} is seen superposed on the profile of the \ion{Mg}{2}
emission line.}
\end{figure}

\begin{figure}
\plotone{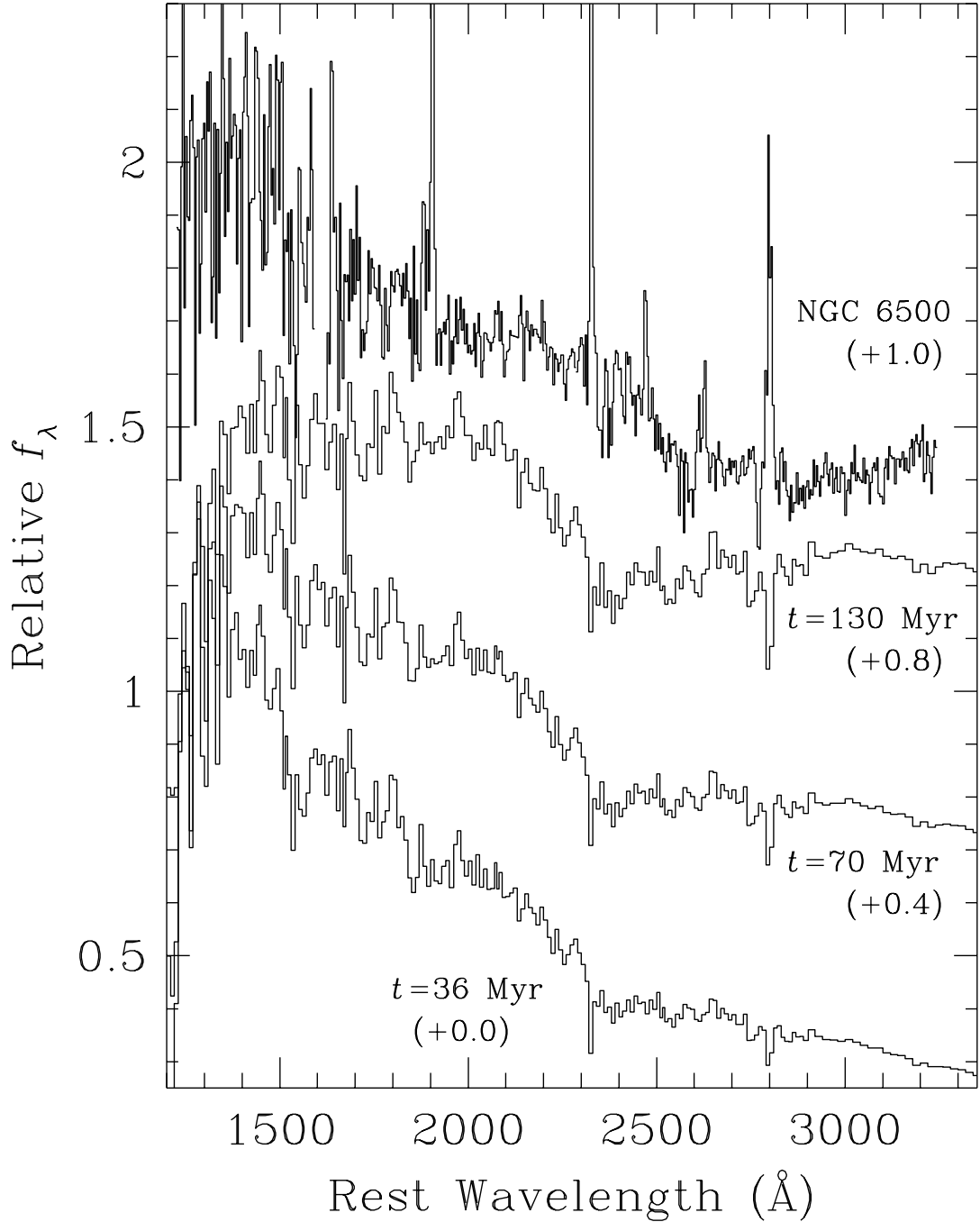}
\caption{Comparison of the UV continuum shape of NGC 6500
with synthetic spectra of single-burst evolutionary models, for ages
of 36, 70, and 130 Myr.  The NGC 6500 continuum has been corrected for
a Galactic extinction of $E(B-V)=0.09$ mag, and the spectrum has been
binned for clarity to 2 \AA\ bin$^{-1}$ above 1600 \AA\ and to 4 \AA\
bin$^{-1}$ below 1600 \AA.  All spectra have been scaled to a common
flux at 2000 \AA.}
\end{figure}

\begin{figure}
\plotone{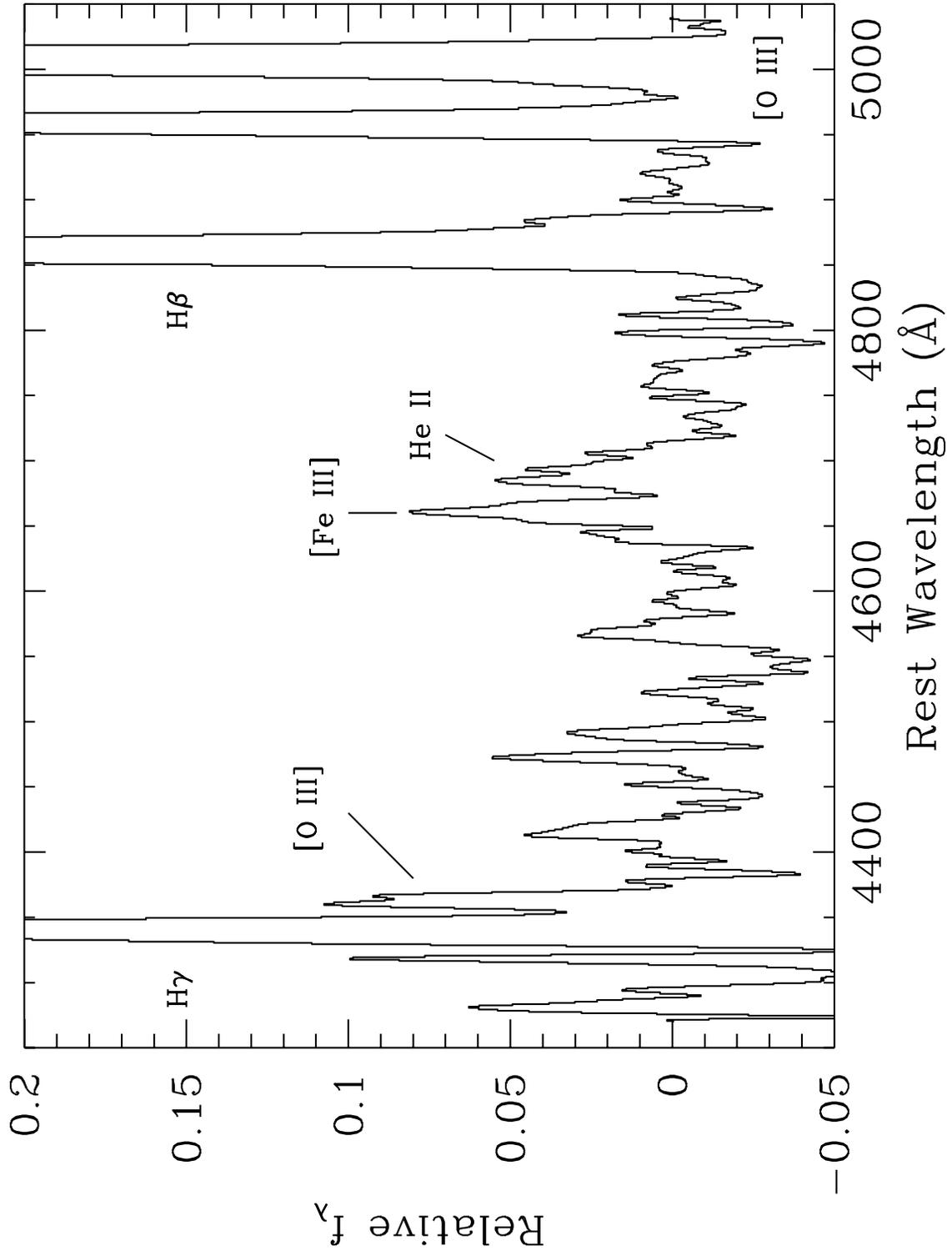}
\caption{Starlight-subtracted optical spectrum of NGC 6500, adapted
from Ho \etal (1997c).  A ``bump,'' part of which may originate from
Wolf-Rayet stars, is present at 4600--4700 \AA.  }
\end{figure}

\begin{figure}
\plotone{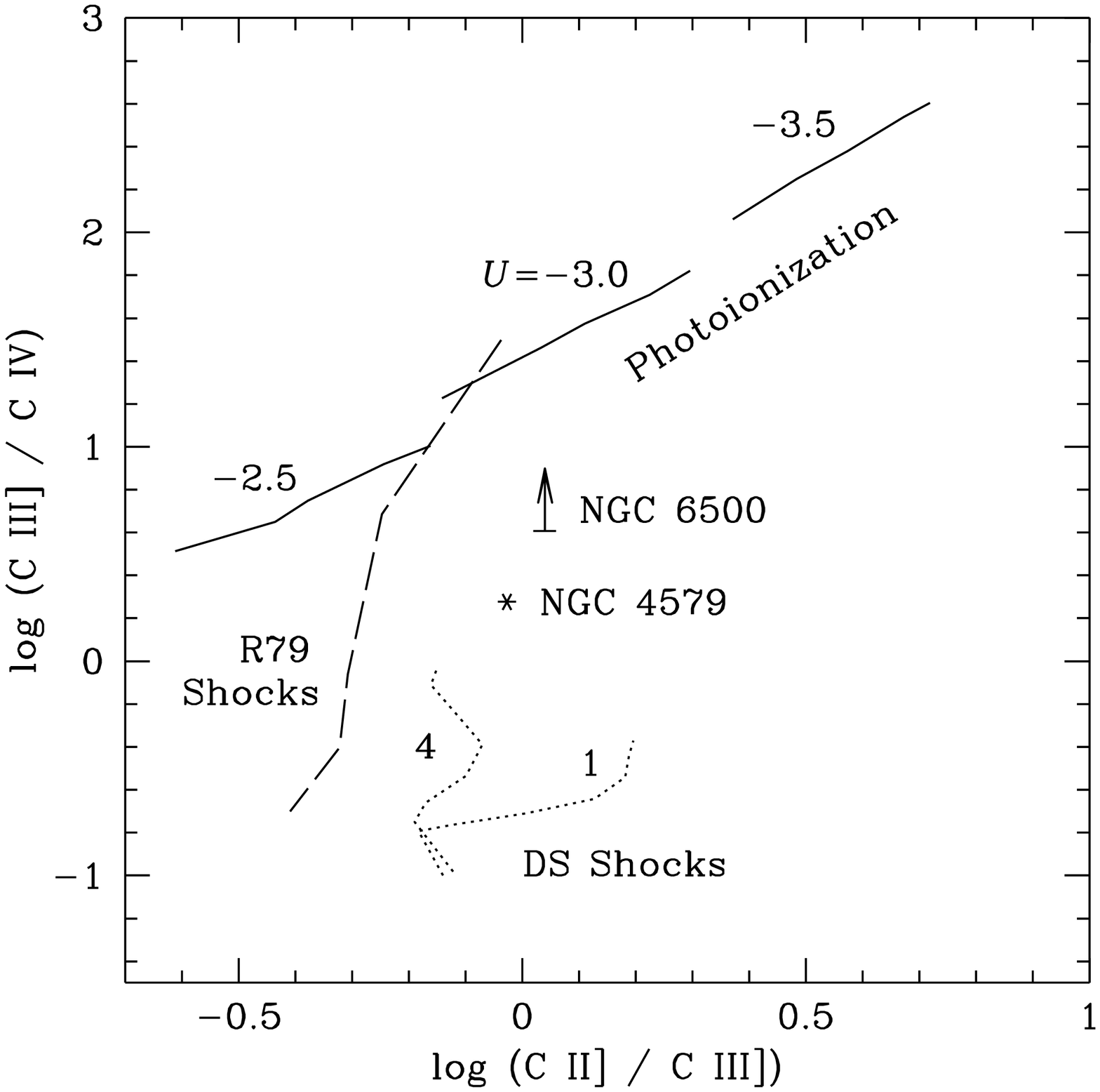}
\caption{Diagnostic diagram for UV emission lines of
carbon: \cii, \ciii, and \civ.  {\it Solid lines:} Power-law
photoionization models computed by Ho \etal (1993), using CLOUDY, for
$f_{\nu} \propto \nu^{-1.5}$ and $\log U = -2.5, -3.0$, and $-3.5$.
Along each curve, density increases from $n_e = 10^{2.5}$ cm$^{-3}$ to
$10^{6}$ cm$^{-3}$, with highest densities at the lower left of each
curve.  {\it Dotted lines:} Shock models of Dopita \& Sutherland
(1996), for magnetic parameter $B/n^{1/2} = 1$ and 4 $\mu$G
cm$^{3/2}$.  Shock velocity ranges from 150 to 500 km s$^{-1}$, with
fastest shocks at the top of each curve.  {\it Long-dashed line:}
Shock models C through G from Raymond (1979), with velocities ranging
from 70.7 km s$^{-1}$ (top) to 141 km s$^{-1}$ (bottom).  For NGC
4579, only the narrow-line component is included.}
\end{figure}

\end{document}